\documentclass[a4paper]{VisionStyle}
\usepackage{epsfig}

\begin{document}

\title{{\sc Xmm--newton} GT and AO-1 observations of SHARC galaxy clusters~:\newline
       RX~J1120.1+4318 \& RX~J0256.5+0006}

\author{S.\,Majerowicz\inst{1} \and M.\,Arnaud\inst{1} \and D.\,Lumb\inst{2} \and D.M.\,Neumann\inst{1}}

\institute{CEA/DSM/DAPNIA/SAp, Saclay, l'Orme des Merisiers, B\^{a}t.\,709, 91191 Gif--sur--Yvette, France
\and
           ESTEC, European Space \& Technology Centre, Keperlaan 1, Postbus 1, 2200 AG Noordwijk, The Netherlands}

\maketitle

\begin{abstract}
We present the follow-up of two galaxy clusters from the Bright and Southern SHARC surveys observed with the {\sc Xmm--newton} satellite. We studied the galaxy cluster RX~J1120.1+4318 which seems to be a relaxed cluster at high redshift according to its surface brightness prof\mbox{}ile and to its temperature prof\mbox{}ile which is flat up to 950\,kpc. Its mean temperature is measured at 5.3\,keV. The other galaxy cluster RX~J0256.5+0006 shows two components. We discuss possible scenario.

\end{abstract}

\section{Introduction}

Catalogs of galaxy clusters throughout a large range of redshifts and cluster {\sc x}--ray luminosities are an ideal basis for the test of cosmological parameters (e.g. \cite{smajerowicz-B3:hen00} ; \cite{smajerowicz-B3:bor01}). The Bright and Southern Serendipitous High redshift {\sc Rosat} Cluster (SHARC) surveys provide a sample of clusters over two decades of {\sc x}--ray luminosity (10$^{43}<$L$_{\mathrm{x}}<$10$^{45}$\,erg/s) with redshifts between 0.2 and 0.8 (\cite{smajerowicz-B3:rom00} ; \cite{smajerowicz-B3:col97} ; \cite{smajerowicz-B3:bur97}).

We present here simple analysis results concerning the follow-up of these selected clusters with the {\sc Xmm--newton} satellite (see \cite{smajerowicz-B3:bar02} for a summary of the project). Our analysis is based on the three {\sc epic} cameras, {\sc mos}1{\&}2 and pn, aboard {\sc Xmm--newton}. Throughout the paper, we use a cosmology with  H$_{0}=50$\,km/s/Mpc, $\Omega_{m}=0.3$ and $\Omega_{\Lambda}=0.7$. The error bars are given with a conf\mbox{}idence level of 90\%.

\section{Data Treatment}

\subsection{Event file cleaning}

The cluster RX~J1120.1+4318 was observed for 20\,ks and RX~J0256.5+0006 for 25\,ks with {\sc Xmm--newton}. We discard the data corresponding to high background periods caused by flares. To monitor this kind of background, we extract light curve in the 10--12\,keV energy band for {\sc mos} data and in the 12--14\,keV for the pn. In these energy bands, the effective area of {\sc Xmm--newton} is negligible and the emission is mainly due to particle induced background. We remove time bins in which there are more than 15\,cts/100\,s for {\sc mos} data and 22\,cts/100\,s for the pn data. The remaining exposure times after the flare rejection are listed in table \ref{smajerowicz-B3_tab:texpo}. The flare rejection method is described in detail by \cite{smajerowicz-B3:maj02}.

\begin{table}[!h]
\caption{Remaining exposure times (in units of ks) for RX~J1120.1+4318 and RX~J0256.5+0006 observations after the flare rejection.}\label{smajerowicz-B3_tab:texpo}
\begin{center}
\leavevmode
\footnotesize
\begin{tabular}[h]{cccc}
\hline \\[-5pt]
Observations & {\sc mos}1 & {\sc mos}2 & pn \\
\hline \\[-5pt]
RX~J1120.1+4318 & 17.6 & 17.9 & 14.2 \\
RX~J0256.5+0006 & 10.6 & 10.3 & 7.1 \\
\hline
\end{tabular}
\end{center}
\end{table}

\subsection{Vignetting correction}

To correct for the vignetting effect --- i.e. the variation of the sensitivity of {\sc Xmm--newton} with off-axis angle and with energy --- we use the method which is described in detail by \cite{smajerowicz-B3:arn01}. This method consists in calculating the ratio of on-axis effective area to effective area at the event position on the detector.

\subsection{Background subtraction method}\label{smajerowicz-B3_sec:dst}

The background subtraction is necessary for extended sources like clusters of galaxies. After removing flares, the background components are the high energy particle induced background, non-vignetted by the telescopes, and the cosmic {\sc x}--ray background (hereafter {\sc cxb}) which is vignetted and depends on sky position of the observation.

To remove these two components, we use a blank sky event file which was produced for each {\sc epic} camera\footnote{They can be retrieved from the Vilspa ftp site~: \textsf{ftp://xmm.vilspa.esa.es/pub/ccf/constituents/extras/background}}. We first subtract the blank field from the source. This step allows to subtract the high energy particle induced background. The second step consists in the correction for the {\sc cxb}. This is done by using data in the region outside the cluster emission. We do the first step for this region and the expected residuals are the difference of emission between the local {\sc cxb} and the {\sc cxb} in the blank field. We finally subtract these residuals to the product of the first step.

This technique to remove all background components is described with more details in \cite{smajerowicz-B3:maj01} and \cite{smajerowicz-B3:pra01}.

\section{RX~J1120.1+4318~: a high z relaxed cluster}

\subsection{Spatial analysis}

\begin{figure}[!h]
\begin{center}
\epsfig{file=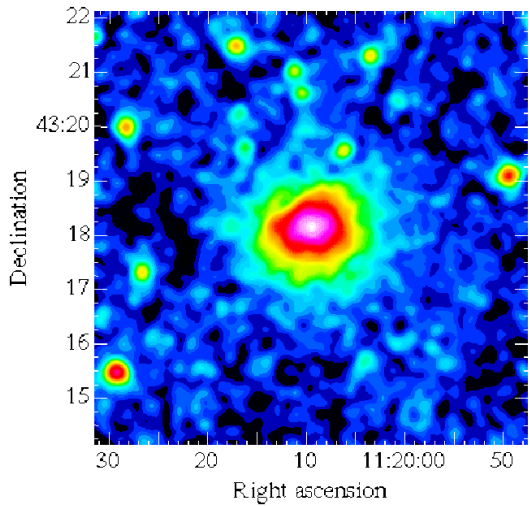, width=7cm}
\end{center}
\caption{Vignetting corrected image of RX~J1120.1+4318 in the 0.3 to 3\,keV energy band f\mbox{}iltered with a Gaussian ($\sigma=5.5$'').}\label{smajerowicz-B3_fig:rxj11img}
\end{figure}

RX~J1120.1+4318 is a distant galaxy cluster at a redshift of 0.6. At this redshift one arc-minute corresponds in our cosmology to 470\,kpc. In the f\mbox{}igure \ref{smajerowicz-B3_fig:rxj11img} where all {\sc epic} cameras were combined, we see that this cluster is most likely in a dynamical relaxed state since its shape of the {\sc x}--ray emission is fairly regular and spherical.

\begin{figure}[!h]
\begin{center}
\epsfig{file=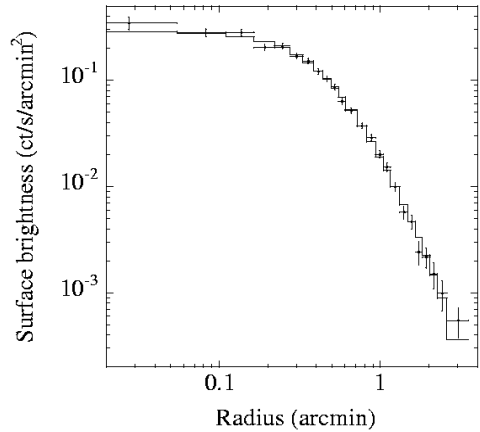, height=6cm, width=7cm}
\end{center}
\caption{Background subtracted surface brightness prof\mbox{}ile of RX~J1120.1+4318 in the 0.3 to 3\,keV energy band from the three {\sc epic} cameras.}\label{smajerowicz-B3_fig:rxj11sbp}
\end{figure}

The surface brightness prof\mbox{}ile of this cluster (see \cite{smajerowicz-B3:maj02} for more explanations about the code) in the energy band between 0.3 and 3\,keV is shown in f\mbox{}igure \ref{smajerowicz-B3_fig:rxj11sbp}. This prof\mbox{}ile is background subtracted and we exclude all point sources in the field of view. We also regroup the data so that we have a signal-to-noise ratio of at least 3\,$\sigma$ in each bin. The cluster is firmly detected up to 3' which corresponds to roughly 1.5\,Mpc.

We f\mbox{}it this prof\mbox{}ile with a $\beta$-model where the surface brightness $S$ is def\mbox{}ined by~:
\begin{equation}
S(r) = S_{0} {(1+(r/r_{c})^{2})}^{-3\beta + 0.5}
\end{equation}
$r_{c}$ is the so-called core radius. We obtain $\beta=0.78^{+0.06}_{-0.04}$ and $r_{c}=0.44^{+0.06}_{-0.04}$\,arc-minute or $207^{+28}_{-19}$\,kpc.

\subsection{Spectral analysis}

\begin{figure}[!h]
\begin{center}
\epsfig{file=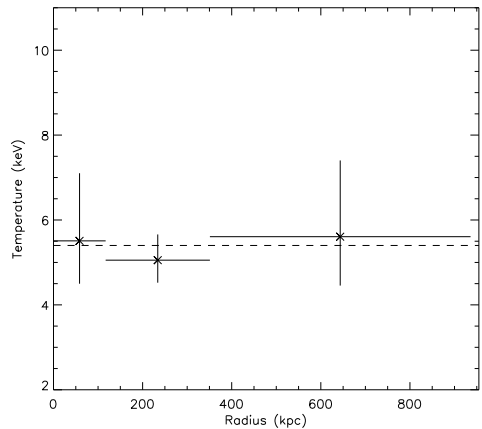, width=7cm}
\end{center}
\caption{Temperature prof\mbox{}ile of RX~J1120.1+4318 (z=0.6). Its mean temperature is displayed with the dashed line.}\label{smajerowicz-B3_fig:rxj11tp}
\end{figure}

{\sc Xmm--newton} observations allow for the first time to determine the temperature profile of a cluster at redshift z=0.6. To derive it, we extract spectra in three concentric annuli and we f\mbox{}it these extracted spectra with an absorbed single temperature plasma model fixing the hydrogen column density to its galactic value.

The measured mean temperature is 5.3$\pm$0.5\,keV with a mean abundance of 0.48$\pm$0.19. The temperature prof\mbox{}ile is flat up to 950\,kpc from the cluster center and does not seem to show a decrease in the center. Furthermore, no excess is observed in the surface brightness profile (see figure \ref{smajerowicz-B3_fig:rxj11sbp}). We can thus conclude that this cluster cannot host a strong cooling flow in its center.

\section{RX~J0256.5+0006~: a merging cluster ?}

\begin{figure}[!h]
\begin{center}
\epsfig{file=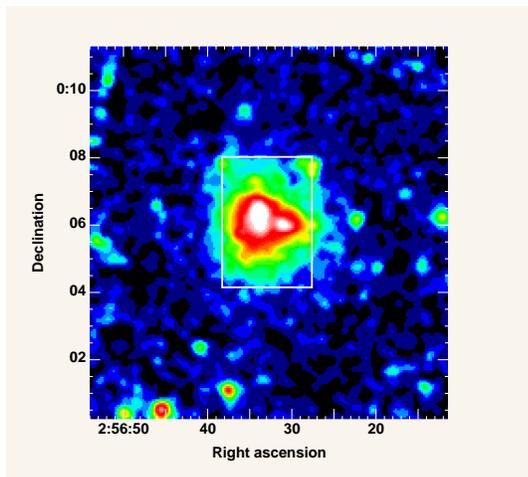, width=7cm}
\end{center}
\caption{Vignetting corrected image of RX~J0256.5+0006 in the 0.3 to 5\,keV energy band smoothed with a Gauss f\mbox{}ilter ($\sigma$=6.6''). The white color box def\mbox{}ines the region where we do a hardness ratio map (see f\mbox{}igure \ref{smajerowicz-B3_fig:rxj02hr}).}\label{smajerowicz-B3_fig:rxj02img}
\end{figure}

RX~J0256.5+0006 is a medium distant cluster at a redshift of 0.36. F\mbox{}igure \ref{smajerowicz-B3_fig:rxj02img} shows cluster image. This cluster appears bimodal. It is of interest to know whether this cluster is in fact in a merger state or whether the bimodal structure is due to chance alignment of two physically unrelated clusters.

\begin{figure}[!h]
\begin{center}
\epsfig{file=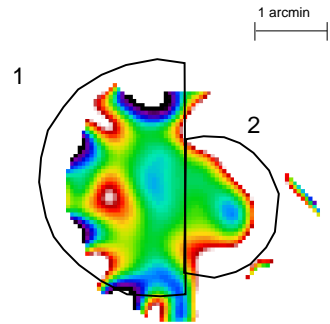, width=6cm}
\end{center}
\caption{Hardness ratio map of RX~J0256.5+0006 between $\sigma=$13.2'' Gauss f\mbox{}iltered and background corrected images in the 0.3--1.3 and 1.3--7.0\,keV energy bands (blue color means cold and white hot).}\label{smajerowicz-B3_fig:rxj02hr}
\end{figure}

Numerical simulations show (\cite{smajerowicz-B3:tak99} ; \cite{smajerowicz-B3:roe97} ; \cite{smajerowicz-B3:sch93}) that cluster mergers create major shock fronts which heat the intracluster medium between the colliding components. To see possible spatial temperature variations in the cluster, we create a hardness ratio map (see f\mbox{}igure \ref{smajerowicz-B3_fig:rxj02hr}). We do not see strong hardness ratio variations between the two components. However this might be due to lack of statistics. But it might also be linked to the possible merger geometry. In case of merging close to the line of sight, the resulting shock waves are widely distributed in the plane-of-sight and not visible for us as concentration between the two components.

Moreover figure \ref{smajerowicz-B3_fig:rxj02hr} shows a hot region which appears in region 1 but the statistics is not suff\mbox{}icient to provide a precise spectroscopic analysis of its temperature.

We extract one spectrum per region def\mbox{}ined in f\mbox{}igure \ref{smajerowicz-B3_fig:rxj02hr} to estimate their mean temperature, and we f\mbox{}ind~: T$_{1}=5.4^{+0.6}_{-0.3}$\,keV and T$_{2}=6.2^{+1.4}_{-1.0}$\,keV.

\begin{figure}[!h]
\begin{center}
\epsfig{file=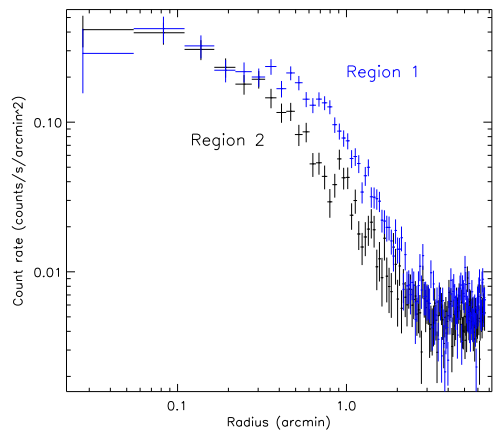, width=7cm}
\end{center}
\caption{Surface brightness prof\mbox{}iles in the 0.3--3\,keV energy band. The selected regions are the SE (region 1) and the NE (region 2) of the main cluster component as def\mbox{}ined by the contours in the lower panel.}\label{smajerowicz-B3_fig:rxj02sbp}
\end{figure}

To know how far are the two components, we let the redshift as a free parameter for the spectrum fits and we f\mbox{}ind z$_{1}=0.34\pm0.02$ and z$_{2}=0.37^{+0.04}_{-0.03}$. The two components can thus be at the same redshift and physically related, or unconnected to each other. In that case it seems that the smallest component is behind the biggest one.

With f\mbox{}igure \ref{smajerowicz-B3_fig:rxj02sbp}, we extract surface brightness prof\mbox{}iles for the biggest component in the region where it is not affected by the emission of the second cluster. From the contours, we can already see the compression of the isophots in region 2. It is clear that the gas distribution is not azimuthally symmetric. This indicates that this component is not in a relaxed state but in a merger state.

In conclusion, we cannot yet definitely conclude on the exact dynamical state of RX~J0256.5+0006, although there are several indications that it is not relaxed. This work is still in progress. In particular, some optical observations will be added and could help us by building galaxy spatial distribution and determining galaxy spectroscopic redshifts.

\end{document}